\documentclass[
aip,
cha,
amssymb
]{revtex4-1}

\usepackage{graphicx}
\def \C{{\mathfrak C}}
\def \N{{\mathbf N}}
\def \R{{\mathbf R}}

{
\noindent {\bf #1 \hfill #2} \\[-2ex]
\begin{flushright}\begin{minipage}{\textwidth}
   }%
{\end{minipage}
\end{flushright}}

\newtheorem{thm}{Theorem}

\newtheorem{lem}[thm]{Lemma}

{{\hfill$\Box$}\endtrivlist}

\begin{document}

\sloppy

\title{The Complexity of Proving Chaoticity and the Church-Turing Thesis}

\author{Cristian S. Calude}
\email{cristian@cs.auckland.ac.nz}
\homepage{http://www.cs.auckland.ac.nz/~cristian}

\affiliation{Department of Computer Science, University of Auckland,
Private Bag 92019, Auckland, New Zealand}

\author{Elena Calude}
\email{e.calude@massey.ac.nz}
\homepage{http://www.massey.ac.nz/~ecalude}
\affiliation{Institute of Information and Mathematical Sciences,
Massey University at Albany,
Private Bag 102-904, North Shore MSC New Zealand}

\author{Karl Svozil}
\email{svozil@tuwien.ac.at}
\homepage{http://tph.tuwien.ac.at/~svozil}
\affiliation{Institute for Theoretical Physics, {\it Vienna} University of Technology,
Wiedner Hauptstrasse 8-10/136, 1040 {\it Vienna},  Austria}

\date{\today}

\begin{abstract}
Proving the chaoticity of some dynamical systems is equivalent to solving the hardest problems in mathematics. Conversely, one argues that it is not unconceivable that classical physical systems may ``compute the hard or even the incomputable'' by measuring observables which correspond to computationally hard or even incomputable problems.
\end{abstract}

\preprint{CDMTCS preprint nr. 384/2010}
\maketitle

\section{Introduction}

Proving that a dynamical system is chaotic is an important problem in chaos
theory~\cite{1985cfd..book.....F}.
Despite causality~\cite{suppes-1993}, {\it virtually any ``interesting'' question about non-trivial dynamical systems appears to be
undecidable}~\cite{Stewart-91}, but  is there a way to mathematically prove {\it this statement?}
Closely related is the question: {\it Is there a way to measure the difficulty of proving the chaoticity of a dynamical system?}
There are only few ``bridges'' between chaotic dynamical systems and
complexity theories, in particular algorithmic information theory~\cite{brudno2,crutchfield1,moore,gacs09}.
The unpredictability of the systems studied in this article comes from a combination
of chaoticity and a ``decision problem'' embedded in the system; the complexity of the  ``decision problem'' (in the sense to be precisely described in the following section)
may be arbitrarily large, including high incomputability.  We shall show that  ``proving the chaoticity of some dynamical systems''
amounts to ``solving the hardest problems in mathematics'' and {\it vice versa.}

We will study a class of mathematical sentences called $\Pi_{1}$--statements. A
sentence of the form $\pi = \forall n \, {\rm Pred} (n)$, where Pred is a computable predicate ($n$ is always a non-negative integer) is called a $\Pi_{1}$--{\em statement}. The Greek letter $\pi$ is used as a generic notation for such a statement; it has no relation with the famous constant $3.145\cdots$. Clearly,  $\pi$ is true if and only if all instances of Pred,
${\rm Pred} (0), {\rm Pred} (1), \dots , {\rm Pred} (n), \dots$ are true.
Every $\Pi_{1}$--statement is {\it finitely refutable} because a single false instance of Pred
makes $\pi$ false. For example, $\forall n [n^{2} + 1 >0]$ is true, but $\forall n [2n+3 \mbox {  is prime}]$ is false.

We deal with formal proofs by using the Zermelo-Fraenkel set theory with the Axiom of Choice (ZFC), the standard system for doing mathematics. So, we say that ``ZFC proves $\pi$'' in case there is a proof in ZFC for $\pi$.

Da Costa,  Doria \cite{dc-d91b} and da Costa,  Doria  and Amaral~\cite{dc-d93} have constructed a two-dimensional Hamiltonian system $\mathcal{H}$
--- a system of first-order differential equations which can be written in the form of Hamilton's equations,
in which the Hamiltonian function represents the total energy of the system ---
with the property that (formally) proving
the existence of a Smale horseshoe in $\mathcal{H}$ is equivalent to (formally) proving Fermat's last theorem. Contrary to the opinion expressed in the above articles, it was  shown that proving that
the two-dimensional Hamiltonian system $\mathcal{H}$ has a Smale horseshoe has low complexity~\cite{ecalude-10} because Fermat's last theorem has a low complexity.

As Fermat's last theorem is a $\Pi_{1}$--statement,  {\it  it is natural to ask whether
the above results can be extended to any $\Pi_{1}$--statement.}
In this note we show that to every $\Pi_{1}$--statement $\pi$
 one can associate a dynamical system $\mathcal{H}_{\pi}$ such that
proving in ZFC the chaoticity of $\mathcal{H}_{\pi}$ is equivalent to proving  $\pi$ in ZFC.  By applying
the computational method~\cite{calude-elena-dinneen06,calude-elena-ec1,calude-elena-ec2}   to  $\Pi_{1}$--statements we
show that there are dynamical systems whose ZFC proofs of their chaoticity are arbitrarily complex and
there are chaotic systems for which ZFC cannot prove their chaoticity.
 The techniques are related to (i) the construction of a Poincar\'e box as a classical physical random number generator
(akin to a quantum Born box), and (ii) the conceivable capability of classical physical systems to
``compute the hard or even the incomputable'' by measuring observables which correspond to computationally hard or even incomputable problems.

\section{$\Pi_{1}$--statements and the complexity measure}

In this section we present a complexity measure~\cite{calude-elena-dinneen06,calude-elena-ec1,calude-elena-ec2}
for  $\Pi_{1}$--statements
%[i.e. statements of the form ``$\forall n \, {\rm Pred}(n)$'',
%where Pred is a computable predicate]
defined by means of register machine programs.

We use a fixed ``universal formalism'' for programs, more precisely, a universal self-delimiting Turing machine $U$. The machine $U$ (which is fully described
below) has to be {\it minimal} in the sense
that none of its instructions can be simulated by a program for $U$ written with the remaining instructions.

To every $\Pi_{1}$--statement $\pi = \forall m {\rm Pred} (m)$ we associate the algorithm $\Pi_{\rm Pred}=\inf\{n\,:\, {\rm Pred}(n) = \mbox{ false}\}$ which systematically searches for a counter-example for $\pi$. There are many programs (for $U$) which implement $\Pi_{\rm Pred}$; without loss of generality, any such program will be denoted also by $\Pi_{\rm Pred}$. Note that
$ \pi \mbox{  is true iff   }  U(\Pi_{\rm Pred})$ never halts.

The complexity (with respect to $U$) of a $\Pi_{1}$--statement $\pi$ is defined by
the length of the smallest-length  program (for  $U$) $\Pi_{\rm Pred}$---defined as above---where minimization is calculated for all possible representations of $\pi$ as $\pi = \forall n {\rm Pred}(n)$:
$C_{U}(\pi) = \min \{ |\Pi_{{\rm Pred}}| \,:\,  \pi = \forall n {\rm Pred}(n)\}$.

For $C_{U}$ it is irrelevant  whether $\pi$ is known to be true or false. In particular, the program containing the  single instruction halt is not a $\Pi_{{\rm Pred}}$ program, for any Pred.  As the exact value
of $C_{U}$ is not important ($C_{U}$ is   incomputable), following a previous article by two of the Authors~\cite{calude-elena-ec2}
we classify $\Pi_{1}$--statements into the following classes:
$\C_{U,n} = \{\pi \,:\,
\pi \mbox{  is a $\Pi_{1}$--statement}, C_{U}(\pi) \le n \mbox{ kbit} \}.$ (Recall that
a kilobit (kbit or kb)
is equal to $2^{10}$ bits.)

We  briefly describe the syntax  and the semantics
of a register machine  language which  implements a (natural) minimal
universal prefix-free binary Turing machine $U$.
Any register program (machine) uses a finite number of registers, each of which may
contain an arbitrarily large non-negative  integer.
By default, all registers, named with a string of lower or upper
case letters, are initialized to 0.  Instructions are labeled
by default with 0,1,2,\ldots

The register machine  instructions are listed below.
Note that in all cases R2 and R3 denote either a register or a non-negative integer, while R1  must be a
register.  When referring to R we use, depending upon the context, either the name of register R or the non-negative integer stored in R.

%\begin{Instruction}{~=R1,R2,R3}{    }
{\bf ~=R1,R2,R3}:
if the contents of R1 and R2 are equal, then the execution continues
at the R3-th instruction of the program;
if the contents of R1 and R2  are not equal, then execution continues with the next instruction
in sequence, and, if the content of R3 is outside the scope of the program,
then we have an illegal branch error.

{\bf ~\&R1,R2}:
the contents of register R1 is replaced by
 R2.

{\bf ~+R1,R2}:
the contents of register R1 is replaced by the sum of the contents of
 R1 and R2.

{\bf ~!R1}: one bit is read into the  register R1, so the contents of R1
becomes either 0 or 1; any attempt to read past the last data-bit
results in a run-time error.

{\bf ~\%}: this is the last instruction for each register machine
program before the input data; it halts the execution in two
possible states: either successfully halts or it halts with an under-read error.

A {\em register machine program}\/ consists of a finite list of
labeled instructions from the above list, with the restriction that
the halt instruction appears only once, as the last instruction
of the list.

\if01
The input data (a binary string) follows immediately
after the halt instruction.  A program not reading the whole
data or attempting to read past the last data-bit results in a
run-time error. Some programs (such as the ones presented in this article)
have no input data; these programs cannot halt with an under-read error.

The instruction {\tt =R,R,n} is used for the unconditional jump to the   $n$-th instruction
of the program.
For Boolean data types we use integers 0 = \verb|false| and 1 = \verb|true|.

For longer programs it is convenient to  distinguish
between the main program and some sets of instructions called ``routines'' which
perform specific tasks for another routine or the main program.  The call and call-back of a routine are executed with  unconditional jumps.
\fi

To compute an upper bound on the complexity of a $\Pi_{1}$--statement $\pi$ we need to compute the size in bits of the program  $\Pi_{\rm \pi}$,
so we need to uniquely code in binary the programs for $U$. To this aim we use a prefix-free coding as follows.

Table~\ref{2010-pi1chaos-t1} enumerates the binary coding of  special characters.
\begin{table}
\caption{Binary encoding of  special characters (instructions and comma); $\varepsilon$ is the empty string.}
\begin{center}
$\begin{array}{ c c| c c}
\hline\hline
{\rm special   } \,\,\, {\rm  characters} & {\rm code} &  {\rm instruction} & {\rm code}\\  \hline
, & \varepsilon  & + & 111 \\
 \& & 01 & ! &  110 \\
= & 00  & \% & 100\\
 \hline\hline \end{array}
$
\end{center}
\label{2010-pi1chaos-t1}
\end{table}
For registers we use the prefix-free regular code ${\rm code}_{1} = \{0^{|x|}1x \mid x\in\{0,1\}^{*}\}$.
%Table~\ref{2010-pi1chaos-t2} contains the codes of the first 14 registers.
The register names are chosen  to  optimize the length of the program, i.e.\
the most frequent registers have the smallest ${\rm code}_{1}$ length.

\if01
\begin{table}
\caption{Binary encoding of the first 14 registers.}
\begin{center}
$\begin{array}{|c|c||c|c|}
\hline\hline
{\rm register} & {\rm code}_{1} & {\rm register} & {\rm code}_{1} \\\hline
{\rm R}_1 & 010 & {\rm R}_8  & 0001001\\
{\rm R}_2 & 011 & {\rm R}_9 & 0001010\\
{\rm R}_3 & 00100 & {\rm R}_{10} & 0001011\\
{\rm R}_4 & 00101 & {\rm R}_{11}  & 0001100\\
{\rm R}_5 & 00110 & {\rm R}_{12} & 0001101\\
{\rm R}_6  & 00111 & {\rm R}_{13} & 0001110\\
{\rm R}_7  & 0001000 & {\rm R}_{14} & 0001111\\
\hline\hline
\end{array}$
\end{center}
\label{2010-pi1chaos-t2}
\end{table}
\fi
%\if01
%\begin{center}
%\[\begin{array}{|c|c||c|c||c|c||c|c|}
%\hline
%{\rm register} & {\rm code}_{1} & {\rm register} & {\rm code}_{1}   & {\rm register} & {\rm code}_{1}   & {\rm register} & {\rm code}_{1} \\\hline
%{\rm R}_1 & 010 & {\rm R}_9 & 0001010 & {\rm R}_{17} &  000010010& {\rm R}_{25} & 000011010\\ \hline
%{\rm R}_2 & 011 & {\rm R}_{10} & 0001011& {\rm R}_{18} &  000010011&{\rm R}_{26} & 000011011\\ \hline
%{\rm R}_3 & 00100 &{\rm R}_{11}  & 0001100 & {\rm R}_{19} & 000010100 & {\rm R}_{27} & 000011100\\ \hline
%{\rm R}_4 & 00101 & {\rm R}_{12} & 0001101& {\rm R}_{20} &  000010101& {\rm R}_{28} &   000011101\\\hline
%{\rm R}_5 & 00110 & {\rm R}_{13} & 0001110 & {\rm R}_{21} &  000010110& {\rm R}_{29} &  000011110\\\hline
%
%{\rm R}_6  & 00111 & {\rm R}_{14} & 0001111 & {\rm R}_{22} &  000010111& {\rm R}_{30} &  000011111\\\hline
%
%{\rm R}_7  & 0001000 &{\rm R}_{15}  & 000010000 &{\rm R}_{23}  &  000011000&
%{\rm R}_{31} & 00000100000\\\hline
%
%{\rm R}_8  & 0001001 &{\rm R}_{16}  & 000010001 & {\rm R}_{24} &  000011001& {\rm R}_{32} & 00000100001\\\hline
%\end{array}\]
%Table~2
%\end{center}
%\fi
For non-negative integers we use the prefix-free regular code  ${\rm code}_{2}
= \{1^{|x|}0x \mid x\in\{0,1\}^{*}\}$.
\if01
Table~\ref{2010-pi1chaos-t3} contains  the codes of
the first 16 non-negative integers.
\begin{table}
\caption{Binary encoding of the first 16 non-negative integers.}
\begin{center}
$\begin{array}{|c|c||c|c||c|c||c|c|}
\hline\hline
{\rm integer} & {\rm code}_{2} & {\rm integer} & {\rm code}_{2}  & {\rm integer} & {\rm code}_{2}  & {\rm integer} & {\rm code}_{2}\\\hline
0 & 100 & 4 & 11010 & 8 &  1110010& 12 & 1110110\\
1 & 101 & 5 & 11011& 9 &  1110011& 13 & 1110111\\
2 & 11000 & 6 & 1110000 & 10 & 1110100 & 14 & 111100000\\
3 & 11001 & 7 & 1110001 & 11 &  1110101& 15 & 111100001
\\\hline\hline\end{array}$
\end{center}
\label{2010-pi1chaos-t3}
\end{table}
\fi
The instructions are coded by self-delimiting binary  strings as follows
(see more details in Refs.~\cite{calude-elena-dinneen06,calude-elena-ec1,calude-elena-ec2}):
\begin{itemize}
\item[{\rm (i)}]  {\tt \&R1,R2} is coded in two different ways, depending on R2 (we omit $\varepsilon$):
$ 01{\rm code}_{1}({\rm R}1){\rm code}_{i}({\rm R}2),$
where $i=1$  if R2 is a register and $i=2$ if R2 is a non-negative integer.

\item[{\rm (ii)}]   {\tt +R1,R2} is coded in two different ways depending on R2:
$111{\rm code}_{1}({\rm R}1){\rm code}_{i}({\rm R}2),$
where $i=1$  if R2 is a register and $i=2$ if R2 is  a non-negative integer.

\item[{\rm (iii)}]  {\tt =R1,R2,R3} is coded in four different ways depending on the data types of  R2 and R3:
$00{\rm code}_{1}({\rm R}1){\rm code}_{i}({\rm R}2){\rm code}_{j}({\rm R}3),$
where $i=1$  if R2 is a register and $i=2$ if R2 is a non-negative integer,
$j=1$  if R3 is a register and $j=2$ if R3 is a non-negative integer.

\item[{\rm (iv)}]  {\tt !R1} is coded by $110{\rm code}_{1}({\rm R1}).$
\item[{\rm (v)}]  {\tt \%} is coded by $100.$
\end{itemize}

For example, Goldbach's conjecture (included in Hilbert's eighth problem~\cite{hilbert-1900e})
states   that {
\it all positive even integers greater than two can be expressed as the
sum of two primes.}
The program  $\Pi_{\rm Goldbach}$  listed in Table~\ref{2010-pi1-t2}
gives the upper bound $C_{U}({\rm Goldbach})
\le 540$ which proves that  the Goldbach conjecture is in the lowest class $\C_{U,1}$.

\if01
\begin{verbatim}
           {\tt        0} {\tt =} {\tt a a 16  }                     {\tt   19} {\tt \&} {\tt c      22    }
           {\tt        1} {\tt \&} {\tt e 2     }                     {\tt   20} {\tt \&} {\tt a      h     }
           {\tt        2} {\tt \&} {\tt d 1     }                     {\tt   21} {\tt =} {\tt a      a 1   }
           {\tt        3} {\tt =} {\tt a e c   }                     {\tt   22} {\tt =} {\tt d      0     }
           {\tt        4} {\tt \&} {\tt d 0     }                     {\tt   23} {\tt \&} {\tt i      0     }
           {\tt        5} {\tt \&} {\tt f e     }                     {\tt   24} {\tt \&} {\tt k      h     }
           {\tt        6} {\tt =} {\tt f a 13  }                     {\tt   25} {\tt =} {\tt k      g 29  }
           {\tt        7} {\tt +} {\tt f 1     }                     {\tt   26} {\tt +} {\tt i      1     }
           {\tt        8} {\tt +} {\tt d 1     }                     {\tt   27} {\tt +} {\tt k      1     }
           {\tt        9} {\tt =} {\tt d e 11  }                     {\tt   28} {\tt =} {\tt a      a 25  }
           {\tt       10} {\tt =} {\tt a a 6   }                     {\tt   29} {\tt \&} {\tt c      32    }
           {\tt       11} {\tt \&} {\tt d 0     }                     {\tt   30} {\tt \&} {\tt a      i     }
           {\tt       12} {\tt =} {\tt a a 6   }                     {\tt   31} {\tt =} {\tt a      a 1   }
           {\tt       13} {\tt =} {\tt d 0 c   }                     {\tt   32} {\tt =} {\tt d      0 35  }
           {\tt       14} {\tt +} {\tt e 1     }                     {\tt   33} {\tt +} {\tt g      2     }
           {\tt       15} {\tt =} {\tt a a 2   }                     {\tt   34} {\tt =} {\tt a      a 17  }
           {\tt       16} {\tt \&} {\tt g 4     }                     {\tt   35} {\tt +} {\tt h      1     }
           {\tt       17} {\tt \&} {\tt h 2     }                     {\tt   36} {\tt =} {\tt a      a 18  }
           {\tt       18} {\tt =} {\tt g h 38  }                     {\tt   37} {\tt \&} {\tt d      0     }
\end{verbatim}
\fi
\begin{table}
\begin{center}
\begin{tabular}{ll|cll|cll|cll}
{\rm 00:\quad \quad}&{\tt  =        a a 16 }   $\quad$&$\quad$&{\rm    11:\quad} &{\tt \& d 0    }  $\quad$&$\quad$&
{\rm    22:\quad} &{\tt = d 0 35   } $\quad$&$\quad$& {\rm 33:\quad} &{\tt +        g 2       }    \\
{\rm 01:}&{\tt \&        e 2    }   &&        {\rm    12:} &{\tt =        a a 6  }  &&       {\rm    23:} &{\tt \&        i 0   } &
&      {\rm 34:} &{\tt =        a a 17    }        \\
{\rm 02:}&{\tt \&        d 1    }   &&        {\rm    13:} &{\tt =        d 0 c  }  &&       {\rm    24:} &{\tt \&        k h   } &
&     {\rm 35:} &{\tt +        h 1       }            \\
{\rm 03:}&{\tt  =        a e c  }   &&        {\rm    14:}&{\tt +        e 1    }  &&        {\rm    25:} &{\tt =        k g  29  } &
&      {\rm 36:} &{\tt =        a a 18    }           \\
{\rm 04:}&{\tt \&        d 0    }   &&        {\rm    15:}&{\tt  =        a a 2  }  &&       {\rm    26:} &{\tt +        i 1   } &
&      {\rm 37:} &{\tt \&        d 0       }       \\
{\rm 05:}&{\tt \&        f e    }   &&        {\rm    16:}&{\tt \&        g 4    }  &&       {\rm    27:} &{\tt +        k 1   } &
&      {\rm 38:} &{\tt   \%                 }        \\
{\rm 06:}&{\tt  =        f a 13 }   &&         {\rm   17:}&{\tt \&        h 2    }  &&       {\rm    28:} &{\tt =        a a  25  } &
&         \\
{\rm 07:}&{\tt  +        f 1    }   &&         {\rm   18:} &{\tt =        g h 38 }  &&       {\rm    29:} &{\tt \&        c 32  } &
&     {\rm  } &{\tt                    }         \\
{\rm 08:}&{\tt  +        d 1    }   &&          {\rm  19:} &{\tt \&        c 22   }  &&      {\rm    30:} &{\tt \&        a i   } &
&     {\rm   } &{\tt                    }         \\
{\rm 09:}&{\tt  =        d e 11 }   &&          {\rm  20:} &{\tt \&        a h    }  &&     {\rm    31:} &{\tt =        a a 1 }  &&     {\rm   } &{\tt                    }         \\
{\rm 10:}&{\tt  =        a a 6  }   &&         {\rm   21:} &{\tt =        a a 1  }  &&       {\rm 32:\quad} &{\tt = d 0 35    }  &
&     {\rm   } &{\tt                       }
\end{tabular}
\end{center}
\caption{\label{2010-pi1-t2} Program $\Pi_{\rm Goldbach}$ for the Goldbach conjecture.}
\end{table}

%\if01
%{\small
%\begin{verbatim}
%0 = a a 16
%1 & e 2
%2 & d 1
%3 = a e c
%4 & d 0
%5 & f e
%6 = f a 13
%7 + f 1
%8 + d 1
%9 = d e 11
%10 = a a 6
%11 & d 0
%12 = a a 6
%13 = d 0 c
%14 + e 1
%15 = a a 2
%16 & g 4
%17 & h 2
%18 = g h 38
%19 & c 22
%20 & a h
%21 = a a 1
%22 = d 0
%23 & i 0
%24 & k h
%25 = k g 29
%26 + i 1
%27 + k 1
%28 = a a 25
%29 & c 32
%30 & a i
%31 = a a 1
%32 = d 0 35
%33 + g 2
%34 = a a 17
%35 + h 1
%36 = a a 18
%37 & d 0
%38 %
%\end{verbatim}
%}
%\fi

\if01
\section{Richardson-Caviness-Wang lemma}
Following Richardson~\cite{richardson68}, Caviness~\cite{321591} and Wang~\cite{wang} we fix a positive integer $n$,
and denote by $\mathcal{E}_{n}$ a set of expressions representing real valued, partially defined functions of real variables and by $F(\mathcal{E}_{n})$
the set of functions represented by the expressions in $\mathcal{E}_{n}$. By $e(x_{1}, x_{2}, \ldots ,x_{n})$ we denote the function  represented by
the expression  $e\in \mathcal{E}_{n}$. Let
$\mu$ and ${\rm sign}$ be the expressions denoting two unary functions such that
$\mu(x) = |x|,\,  {\rm sign}(x) = 1$, for all $x\not= 0$.
We assume that $\mathcal{E}_{n}$
is generated by:
\begin{itemize}
\item[{\rm (i)}]   the rational numbers and $\pi$ as constant functions,
\item[{\rm (ii)}]    variables $x_{1}, x_{2}, \ldots ,x_{n}$,
\item[{\rm (iii)}]     the functions $\sin, \mu, {\rm sign}$, and
\item[{\rm (iv)}]    the operations of  addition, subtraction, multiplication and composition.
\end{itemize}

In fact we can omit subtraction and $\pi$, cf. Ref.~\cite{wang,Laczkovich-2002}.

Let $\mathcal{P}_{n}$ be the set of  polynomials with integral coefficients
in the variables $x_{1}, x_{2}, \ldots ,x_{n}$. Let $\N, \R$ denote the set of non-negative integers and reals, respectively.

\begin{lem}\label{equiv} {\rm Ref.~\cite{richardson68,321591,wang} }
For every polynomial $P(x_{1}, x_{2}, \ldots ,x_{n}) \in \mathcal{P}_{n}$ there exists $f_{P}(x_{1}, x_{2}, \ldots ,x_{n}) \in F(\mathcal{E}_{n})$ such that the following conditions are equivalent.
\begin{itemize}
\item[{\rm (i)}] $\exists x_{1}, x_{2}, \ldots ,x_{n} \in \N$:
$P(x_{1}, x_{2}, \ldots ,x_{n})=0$.
\item[{\rm (ii)}]  $\exists  x_{1}, x_{2}, \ldots ,x_{n} \in \R$:
$f_{P}(x_{1}, x_{2}, \ldots ,x_{n})=0$.
\item[{\rm (iii)}]
$\exists  x_{1}, x_{2}, \ldots ,x_{n} \in \R$:
$f_{P}(x_{1}, x_{2}, \ldots ,x_{n})\le 1$.
\end{itemize}

\end{lem}
\fi

\section{Main results}

We start with a result relating $\Pi_{1}$--statements and Hamiltonians.

\begin{thm} \label{CC}Assume {\rm ZFC} is arithmetically
sound, i.e.\ every statement {\rm ZFC} proves is true. Then, to each $\Pi_{1}$--statement
$\pi= \forall m \, {\rm Pred}(m)$
one can effectively construct in the formal language of {\rm ZFC}
a  Hamiltonian system $\mathcal{H}_{\pi}$ such that  {\rm ZFC} proves that the system
$\mathcal{H}_{\pi}$  has  a Smale horseshoe  iff   {\rm ZFC} proves $\pi$.
\end{thm}

We denote by $h$ and $k$
 the Hamiltonian for the two-dimensional
system with a Smale horseshoe as defined by Holmes and Marsden~\cite{Homes-Marsden-82}  (their Example~4)
and the Hamiltonian for the free particle, respectively.  Clearly, the systems  $h$ and $k$ can be represented in the formal language of ZFC.  Define
the  Hamiltonian $ \mathcal{H}_{\pi}^{m}$  as a linear combination of $h,k$:

\begin{eqnarray}
\label{H}
 \mathcal{H}_{\pi}^{m} (q_{1}, \dots ,q_{n}, p_{1}, \dots ,p_{n}) & = &  {\rm Pred}(m) \cdot h (q_{1}, \dots ,q_{n}, p_{1}, \dots ,p_{n})\\
 &  & + (1- {\rm Pred}(m))\cdot k(q_{1}, \dots ,q_{n}, p_{1}, \dots ,p_{n}).\nonumber
\end{eqnarray}

Fix a positive integer $i$. In view of (\ref{H}), $\mathcal{H}_{\pi}^{i}$ can be represented in the formal language of ZFC and $\mathcal{H}_{\pi}^{i} (q_{1}, \dots ,q_{n}, p_{1}, \dots ,p_{n})$ $ =
h(q_{1}, \dots ,q_{n}, p_{1}, \dots ,p_{n})$  iff $\mbox{  ZFC proves  }  \pi $. In case the above equivalence holds true, $\mathcal{H}_{\pi}^{i} (q_{1}, \dots ,q_{n}, p_{1}, \dots ,p_{n}) =
\mathcal{H}_{\pi}^{j} (q_{1}, \dots ,q_{n}, p_{1}, \dots ,p_{n})$, for all  non-negative integers $i, j$, hence we can name each $\mathcal{H}_{\pi}^{t}$  by $\mathcal{H}_{\pi}$.

We have shown that \\[-5ex]
\begin{quote}
 $\mbox{  ZFC proves  }  \pi \mbox{  iff  ZFC proves  that } \mathcal{H}_{\pi} \mbox{   has a Smale horseshoe}$,
 \end{quote}
hence  ending the proof of Theorem~\ref{CC}.

%\end{proof}

If $\pi$ is true but unprovable in ZFC, then the equality $ \mathcal{H}_{\pi}^{i} (q_{1}, \dots ,q_{n}, p_{1}, \dots ,p_{n})$ $ =
h(q_{1}, \dots ,q_{n}, p_{1}, \dots ,p_{n})$ is true but unprovable in ZFC.

In case $\pi$ is the Fermat's last theorem, Theorem~\ref{CC} is exactly the result
proved~\cite{dc-d91b,dc-d93};  our direct proof does not need the machinery involving Richardson lemma used in Ref.~\cite{dc-d91b,dc-d93}.

Theorem~\ref{CC} can be applied to a variety of $\Pi_{1}$--statements including Goldbach's conjecture,
  Riemann's hypothesis, the four color theorem, and many others.

\medskip

We address now the complexity issue:
How difficult is it
to prove  in ZFC that the system $\mathcal{H}_{\pi}$ in Equation~(\ref{H}) is chaotic?
Using the
complexity $C_{U}$ we can show that Fermat's last theorem and Goldbach's conjecture are in $\C_{U,1}$,   the Riemann hypothesis is in
   $\C_{U,3}$, and the four color theorem is in $\C_{U,4}$~\cite{calude-elena-ec2,calude-elena-CE3,ecalude-09}; their corresponding
   dynamical systems produced by Theorem~\ref{CC} have the property that the complexity of its chaoticity proof is in the corresponding class.

As for every natural $n$ there exists a natural $m_{n}$ such that $\C_{U,n} \subset \C_{U,m_{n}}$, it follows that,
according to $C_{U}$, there exist arbitrarily complex $\Pi_{1}$--statements; hence
proving the chaoticity
of the  system $\mathcal{H}_{\pi}$ can be arbitrarily complex.

Finally, there are infinitely many
true, but unprovable in ZFC, $\Pi_{1}$--statements $\pi$~\cite{cal-paun-83},
such that
the corresponding systems $\mathcal{H}_{\pi}^{i}$ are chaotic but ZFC cannot prove their chaoticity.
For example, from the negation of the halting problem for $U$ we get infinitely many $\Pi_{1}$-statements $\pi_{x} =$ ``$\forall n$
($U(x)$ does not stop in time $n$)'' which are undecidable in ZFC.

\section{Computational capabilities of chaotic motion}

One of the intriguing possibilities of the aforementioned equivalences between certain statements in ZFC and chaotic motion is
the hypothetical possibility to ``decide'' hard problems in ZFC
or ``perform incomputable tasks'' by observing the corresponding chaos~\cite{Scarpellini-63,Stewart-91,dc-d93,Scarpellini-2003,Scarpellini-2003c}.
Indeed, if such methods and procedures have an ``effective'' physical implementation,
then, strictly speaking, the Church-Turing thesis identifying the informal notion of {\em computable algorithm}
with {\em Turing computability}, or, equivalently, {\em recursive functions}, is too restricted and has to be adapted to the
physical capacities~\cite{davis-58,rogers1,kreisel} (for a converse  viewpoint restricting operations to strictly finitistic means,
see Refs.~\cite{bridgman,gandy2,gandy1}).

It is rather intriguing that, at least in this respect, the situation resembles the famous Einstein, Podolski and Rosen (EPR) argument~\cite{epr} for
a possible ``incompleteness'' of quantum mechanics.
According to EPR, whereas quantum theory does not allow complementary physical observables to simultaneously ``exist,''
experiment (augmented with counterfactual reasoning) allows for such ``elements of physical reality.''

In the case of chaotic systems, our present theory of computability,
formalized by recursion theory, does not allow the ``execution'' of certain ``hard'' tasks;
but the equivalent chaotic systems would perform just such tasks, sometimes with relative ease on the side of the experimenter.
One example of such seemingly mismatch --- in the sense of EPR --- of computability theory and physical computation is the construction of
``oracles producing random bits,'' as discussed in the next section.

\section{Poincar\'e box as physical random number generator}

Chaotic systems can be used as a physical device for incomputability.
In the ``extreme'' algorithmically incompressible case, a chaotic dynamical system can serve
as a source of random bits; i.e., as a physical {\em random number generator} (RNG).
This RNG can be conceptualized by enclosing a chaotic system in a ``black box'' with an output interface
which communicates the consecutive physical states of the chaotic evolution~\cite{PhysRevLett.45.712} in a properly encoded symbolic form.
In order for these, say, strings of bits, to be physically certified random, it is necessary to ascertain chaoticity;
a property which relates to the proofs of chaoticity discussed above.

This scenario can be elucidated by considering the shift map $\sigma$
(a form of generalized shift studied by Moore~\cite{moore})  which
``pushes'' up successive bits of the sequence $s=0.s_1s_2s_3\cdots$;
i.e., $\sigma (s)= 0.s_2s_3s_4\cdots$,  $\sigma (\sigma (s))= 0.s_3s_4s_5\cdots$, and so on.
Suppose one starts with an initial ``measurement'' precision of, say, just one bit after the comma,
indicated by a  ``window of measurability;''
all other information ``beyond the first bit after the comma'' is hidden to the experimenter at this point.
Consider an initial state represented by an algorithmically random real $s$.
At first the experimenter records the first position $s_1$ of $s$, symbolized by
$0.[[s_1]]s_2 s_3\cdots$, where the square brackets ``$[[~\cdots~]]$''
indicate the boundaries of the experimenter's sliding ``window of measurability.''
Successive iterations of the shift map ``bring up'' more and more bits of the initial sequence of $s$; i.e.,
$\sigma (s)$ yields $0.s_1[[s_2]]s_3s_4\cdots$,
$\sigma (\sigma (s))$ yields $0.s_1 s_2[[s_3]]s_4s_5\cdots$, and in general
$\sigma^{(i)} (s)$ yields $ 0.\cdots s_{i-1}s_{i}[[s_{i+1}]]s_{i+2}s_{i+3}\cdots$ after $i$ iterations of the shift map.
Thus effectively, the algorithmic information content of $s$ ``unfolds'' at a rate of one bit per time cycle.
If $s$ is algorithmically random, then (at least ideally) the empirical recording of its successive bits generates a random sequence (in the asymptotic limit).

It is not totally unreasonable to conjecture that, with respect to algorithmic (hence also statistical)  tests of randomness,
{\em Poincar\'e boxes} {\em cannot} be differentiated from another type of physical RNGs termed {\em Born boxes}, which are based on quantum indeterminism
(e.g., photons impinging on beam splitters and detectors~\cite{svozil-qct,rarity-94,zeilinger:qct,stefanov-2000,0256-307X-21-10-027,wang:056107,fiorentino:032334,svozil-2009-howto}).
Considering the different physical origins of physical indeterminism exploited by the Poincar\'e and Born boxes
--- in the first, classical case, indeterminism resides in the continuum, whereas in the second, quantum case,
in the postulated~\cite{born-26-1,born-26-2,zeil-05_nature_ofQuantum,2008-cal-svo} irreducible randomness of certain individual outcomes involving photons ---
why should the two physical RNG's perform equally from an algorithmic information theoretic~\cite{chaitin3,calude:02} point of view?
Because, one could argue,
both would produce (in the asymptotic regime) random strings with high probability.

The Poincare box derives its random behavior from a {\em single, individual} initial value containing incompressible algorithmic information with probability
one~\cite{brudno2,crutchfield1},  whereas
the Born box utilizes {\em successive, independent} ideal coin tosses.
Whether or not these speculations are justified or not only experiment can tell.
So far, no empirical evidence either for or against the conjectured equivalence of Poincar\'e and Born boxes exist.

It is not too difficult to ``construct'' a Poincar\'e box by utilizing a shift map which ``pumps'' up
the bits of the binary representation of the initial value by one bit per (discrete iteration) cycle.
Of course, assuring the physical representability of this extreme chaotic regime for concrete
classical chaotic systems, might turn out to be a ``hard'' task; as has been argued above.
With this proviso, and by further assuming that the initial value is some element of the continuum
(in ZFC the ``selection'' of an initial value is guaranteed by the Axiom of Choice), the shift map
is, at least asymptotically, capable of yielding a random number with probability one.

\section{Summary and outlook}

We have argued that every $\Pi_{1}$--statement $\pi$ can be associated with a dynamical system $\mathcal{H}_{\pi}$ such that
ZFC proves the chaoticity of $\mathcal{H}_{\pi}$ iff ZFC proves $\pi$.
Many ``hard''problems, such as, for example, the Riemann hypothesis and the four color theorem, are  $\Pi_{1}$--statements.
The computational method~\cite{calude-elena-dinneen06,calude-elena-ec1,calude-elena-ec2}  has been applied to
$\Pi_{1}$--statements, resulting in a complexity measure for proving the chaoticity of some dynamical systems.
Consequently, there are dynamical systems for which the ZFC proofs of their chaoticity are
arbitrarily complex according to the above complexity measure.
Furthermore, there are infinitely
many chaotic systems for which ZFC cannot prove their chaoticity.

One of the challenging conceptual questions which is motivated by these results is the issue of relating physical entities to formal ones.
In particular at stake is the Church-Turing thesis, which is challenged from a classical physical perspective.
As classical chaotic motion seems to be capable to ``perform'' incomputable tasks --- a criterion which might,
as we argue, be ``hard'' to certify for a wide variety of Hamiltonian systems, but which nevertheless is a feasible scenario  ---
it might not be too unreasonable to speculate that
the present formal theories of computability would have to be adapted in accordance with our physical capabilities originating
from chaotic motion.

\section*{Acknowledgement} We thank Alastair Abbott and the anonymous referees for critical comments which improved the article.

%\bibliography{svozil}

%merlin.mbs 2010-03-15 4.21a (PWD, AO, DPC)
\end{document}